\documentclass[useAMS,usenatbib,aas_macros]{mn2e}
\usepackage{natbib}
\usepackage{amsmath}
\usepackage{amssymb}
\usepackage{amstext}
\usepackage{graphicx}
\usepackage{epsfig}
\usepackage{verbatim} 
\usepackage{aas_macros}

\title[Feedback from X-ray binaries]{Another thread in the tapestry of stellar feedback: X-ray binaries}

\author[Justham \& Schawinski]{Stephen Justham$^{1,2}$ \& Kevin Schawinski$^3$\thanks{Einstein Fellow} \\
$^1$National Astronomical Observatories, The Chinese Academy of Sciences, Beijing 100012, China; sjustham@bao.ac.cn \\ 
$^2$The Kavli Institute for Astronomy and Astrophysics, Peking University, Beijing 100871, China \\
$^3$Department of Physics \& Yale Center for Astronomy and Astrophysics, Yale University, CT 06520-8121, U.S.A.; kevin.schawinski@yale.edu}

\date{Accepted 2012 March 23. Received 2012 March 16; in original form 2011 April 27.}

\begin{document}

\label{firstpage}
\maketitle

\begin{abstract}
We consider X-ray binaries (XBs) as potential sources of stellar feedback.  XBs observationally appear able to deposit a high fraction of their power output into their local interstellar medium, which may make them a non-negligible source of energy input.  The formation rate of the most luminous XBs rises with decreasing metallicity, which should increase their significance during galaxy formation in the early Universe.  We also argue that stochastic effects are important to XB feedback (XBF) and may dominate the systematic changes due to metallicity in many cases. Large stochastic variation in the magnitude of XBF at low absolute star formation rates provides a natural reason for diversity in the evolution of dwarf galaxies which were initially almost identical, with several percent of such halos experiencing energy input from XBs roughly two orders of magnitude above the most likely value.  These probability distributions suggest that the effect of XBF is most commonly significant for total stellar masses between approximately $10^{7}$ and $\rm 10^{8}~M_{\odot}$, which might resolve a current problem with modelling populations of such galaxies.   
We explain how XBs might inject energy before luminous supernovae (SNe) contribute significantly to feedback and how XBs can assist in keeping gas hot long after the last core-collapse SN has exploded. Energy input from XBs produces different behaviour to that from SNe, partly since the peak energy input from a mean XB population continues for $\approx 100$ Myr after the start of a starburst.  XBF could be especially important to some dwarf galaxies, potentially heating gas without expelling it; the properties of XBF also match those previously derived as allowing episodic star formation. We also argue that the efficiency of SN feedback (SNF) might be reduced when XBF has had the opportunity to act first.  In addition, we note that the effect of SNF is unlikely to be scale-free; galaxies smaller than $\approx 100$ pc might well experience less effective SNF.
\end{abstract}


\section{Introduction}
\label{sec:intro}

A collection of processes jointly known as ``feedback'' seem to be vital in governing the state of the baryonic content of dark matter haloes; understanding these processes is therefore extremely important to the study of galaxy evolution and structure formation.  The current-day baryon content of an $L_{*}$ galaxy's halo seems to be lower than the cosmic mean, suggesting that baryons have been ejected from them, and feedback appears to have an even greater effect on the lowest-mass and highest-mass galaxies, truncating the high-mass end of the galaxy mass function and flattening the low-mass end \citep[see, e.g.,][]{Kauffmann+1999, Benson+2003, Cole+2001, Tegmark+2004, Croton+2006, Bower+2006}.

Although it is agreed that feedback mechanisms are required to produce the properties of galaxies in the observed Universe, the details of feedback remain far from resolved.   It is widely accepted that stellar feedback is dominant in low-mass galaxies, the canonical sources of which are supernovae (SN) and irradiation and winds from massive stars. The long-known uncertainty in SN feedback (SNF) has been the efficiency with which supernova explosions can couple to the interstellar medium \citep[see, e.g.,][]{Larson1974, Chevalier1977, McCraySnow1979, Spitzer1990}. Canonically, only a tenth of a typical supernova's kinetic explosion energy is available for feedback, since the rest is radiated away \citep{Larson1974}, but the debate about the underlying physics and characteristics of SNF is ongoing.
For example, \citet{CeverinoKlypin2009} argued that SN feedback may be dominated by the supernovae from runaway stars which are able to explode in cold gas, since SNe exploding in hot gas lose more of their energy to radiation. The fact that around 40 per cent of luminous core-collapse supernovae seem to occur in runaway stars, as deduced from observations \citep{JamesAnderson2006} and population synthesis \citep{Eldridge+2011} supports the potential significance of this effect.

Assuming that SNF is dominant, the classic work of \citet{DekelSilk1986} found that `The critical condition for global gas loss as a result of the first burst of star formation is that the virial velocity be below a critical value on the order of 100 km $\rm s^{-1}$'. However, some of the lowest-mass dwarfs (with virial velocities $\approx$10 $\rm km ~s^{-1}$) appear to somehow retain gas after the first episode of star formation \citep[see, e.g., ][]{Irwin+2007}; hence in some cases feedback seems remarkably weak. Yet feedback may often need to be strong in order to match the observed dwarf galaxy population with the predictions of CDM \citep[see, e.g.,][]{Koposov+2009}. Continuing inadequacies in our understanding of stellar feedback is one of the possible explanations suggested by \cite{Sawala+2011} for the fact that models of galaxy formation and evolution in $\approx 10^{10} {\rm ~M_{\odot}}$ halos produce results apparently inconsistent with observations. For such reasons we must continue to examine all possible feedback mechanisms.

In this paper, we argue that another source of stellar feedback -- X-ray binaries (XBs) -- should be included in galaxy modelling \citep[see also][]{Cantalupo2010}.  X-ray binary feedback (XBF) should be especially important at low metallicity and hence in the early epochs of galaxy formation \citep[see also the independent work of][]{Mirabel+2011}.\footnote{\citet{Power+2009} previously considered the contribution of XB photoionisation to reionisation; see the \textit{Note Added In Proof}.} 
We emphasise how the properties of XBF seem qualitatively different to those of SN feedback, and that these differences are interesting and important. We also argue that the stochastic variation in XBF should be expected to be more significant than for SNF. Partly for that reason, XBF seems promising in helping to explain the complex star-formation histories and diversity of very low-mass dwarf galaxies. We also note that strong XBF can begin before strong SNF, which may allow XBF to modify the effects of SNF.

The class of XBs contains objects with a wide range of ages and luminosities. Accreting stellar-mass black holes (BHs) and neutron stars were once broadly divided into two groups; low-mass X-ray binaries (LMXBs) where mass is transferred from the low-mass ($\lesssim 2 \rm~M_{\odot}$ donor star) by Roche-lobe overflow and high-mass X-ray binaries (HMXBs) where the high-mass ($\gtrsim$10$\rm~M_{\odot}$) donor star spontaneously loses mass in a wind or excretion disc. The majority of luminous systems known as ultra-luminous X-ray sources (ULXs) are probably XBs, but in neither of those old categories; they are expected to contain intermediate-mass or massive donor stars which are undergoing short-lived phases of Roche-lobe overflow (see, e.g., \citealt{Gilfanov+2004SFR}; \citealt{RPP2005}, hereafter referred to as RPP; \citealt{Madhu+2006};  \citealt{Blecha+2006}; \citealt{Zezas+2007}).   The nature of ULXs does not affect our main conclusions, which are consistent with empirical studies of XB populations, as we describe later. However, some of the details in this paper do rely on models which treat ULXs as stellar-mass BHs.  We mostly consider populations of XBs rather than individual sources, and in young stellar populations then XBs with more massive donor stars (including, we expect, ULXs) should dominate the luminosity of the XB population; most of this work discusses the potential effects of feedback from these sources.  In old stellar populations then LMXBs are still present, and we also briefly consider to what extent LMXBs might keep gas hot in non-star forming galaxies.

The importance of binary stars on the appearance of stellar populations, and therefore on the appearance of galaxies, has been demonstrated before \citep[see, e.g.,][]{Vanbeveren+1997,Vanbeveren+1998,Belkus+2003,Zhang+2005,DionneRobert2006,Han+2007,EldridgeStanway2009}. The potential effect on galactic structure of a subset of compact binaries other than XBs has also recently been considered by \citet{MooreBildsten2011}; binary stars matter on galactic scales.

Section \ref{sec:motivating} outlines some of the reasons why we consider that energy input from XBF could be important in shaping galaxies. This includes a brief summary of the observed effects of some XBs on the ISM, the likely metallicity dependence of the formation rate of young XBs, and how XBF can operate \emph{before} SNF. There we also outline why we think that XBF may be particularly relevant to some of the observed characteristics of dwarf galaxies. Section \ref{sec:XBenergy} expresses in more detail the energy available from mean populations of XBs, and provides expressions for use in semi-analytic galaxy models.   
In Section \ref{sec:fluctuations} we examine potential stochastic variations in XBF, and suggest that they provide a natural reason for diversity in the evolution of initially identical dwarf galaxies.  In particular, Section \ref{sec:ULXflucts} finds that the most important factor in determining the effect of XBF on \emph{individual} low-mass halos may not be metallicity but probability.  Furthermore, in Section \ref{sec:peakdominance} we argue that natural statistical variation in XBF may explain why the effect of XBF would be most commonly significant in populations of galaxies with stellar mass approximately between $10^{7}$ and $\rm 10^{8}~M_{\odot}$, potentially resolving the problems with current galaxy evolution models which were identified by \cite{Sawala+2011}.  Our discussion speculates on how the interaction between XBF and SNF may lead to a rich variety of behaviour.

\section{Motivating X-ray Binary Feedback}
\label{sec:motivating}

\subsection{Efficiency of energy coupling \& demonstrable outflows}
\label{sec:efficency}

A large fraction of the energy output of at least some XBs seems to be deposited into the ISM. By using ionisation nebulae around ULXs as calorimeters, \citet{PakullMirioni2003} found that the energy deposited in the nebulae was comparable to the observed X-ray luminosity \citep[assuming isotropic X-ray emission; see also ULX IC 342 X-1,][]{Roberts+2003}.   The potential photoionisation effect of XBs on galactic-scale stellar feedback has been independently considered by \citet{Cantalupo2010} and \citet{Mirabel+2011}.$^{1}$

Photoionisation is not the only potential route through which XBs might affect their environment.  Other XBs also seem efficient in depositing energy in the ISM to produce kinetic outflows: \cite{Gallo+2005} found that  {\it`the total power dissipated by the jets of Cygnus X-1 in the form of kinetic energy can be as high as the bolometric X-ray luminosity of the system' }. The nebula around SS433 (W50) appears to be jet-inflated, and SS433 drives outflows \citep[see, e.g.,][]{Blundell+2001, Goodall+2011}. The size of the XB-inflated nebula observed by \citet{Pakull+2010} is $\approx 300 {\rm~pc}$. This kinetic input from XBs was considered by \citet{Fender+2005} as a potential driver for cosmic-ray emission on galaxtic scales.

Indeed, some XBs put \emph{far more} energy into local outflows than into their X-ray luminosity: \citet{Pakull+2010} argue that the kinetic output of one microquasar is $\sim 10^{4}$ times its X-ray power. This XB is \emph{not} a ULX, but its kinetic energy output is a few times $\rm 10^{40}~erg~s^{-1}$ \citep[see also][]{Soria+2010}. If such efficient deposition of energy into the ISM is common, then we vastly underestimate the magnitude of XBF in what follows.  In that case then the majority of the feedback would be in kinetic outflows rather than photoionisation.

\subsection{Metallicity dependence}
\label{sec:metallicityintro}

The abundance of bright, young BH XBs (in particular ULXs) is expected to increase at lower metallicity ($Z$). \citet{Linden+2010} and Justham \& Podsiadlowski (in prep.) have studied the metallicity-dependence of luminous BH XBs, roughly agreeing that the formation rate at $Z \approx 0.01~{\rm Z_{\odot}}$ could be $\sim$10 times higher than at $\rm Z_{\odot}$. Explaining the detailed reasons for this metallicity dependence is left to those papers, but it is not enough to argue that BH formation from single stars should be more common at low metallicity; the changing formation rate of interacting binaries containing BHs as a function of metallicity turns out to be dominated by changes in the available range of initial binary separations which can go on to produce such systems. Specifically, the main effect relates to the allowed timing of the interaction when the envelope of the BH-forming star is removed (with respect to the nuclear evolution of that star).  For non-conservative mass transfer, \citet{Dray2006} found a BH HMXB formation rate at $Z=0.001$ that is $\approx$10 times higher than at $Z=0.02$.  Earlier work by \citet{Belczynski+2004} had also found that the number of young binaries consisting of a BH with a main-sequence star (BH+MS) should change by a large factor as a function of metallicity. Belczynski et al.\ looked at model populations with metallicities of $Z=10^{-4}, 10^{-3}$ and $0.02$. At an age of 11 Myr, the ratio of BH+MS binaries between these models was  $\approx$  9 : 5.7 : 1, and at 104 Myr, $\approx$ 6.7 : 4.5 : 1.  The production rate of BH+MS systems is not the same as that of ULXs, but it should be indicative of the production rate of bright BH XBs.

All the XB formation calculations discussed in this subsection have assumed that the low-metallicity IMF is the same as the present-day IMF, so a top-heavy IMF at low metallicity would enhance this metallicity dependence. (We note in passing that binaries themselves should affect observational determinations of the IMF: see, e.g., \citealt{EldridgeIMF2011}.)

Furthermore, the typical masses of stellar BHs may well be higher at lower metallicity \citep[see, e..g.,][]{EldridgeTout2004}.  Any increase in the typical mass of \emph{accreting} BHs at lower metallicity should increase the potential luminosity of the XB population further (see, e.g., \citealt{Linden+2010}).  We note that a novel formation mechanism has recently been proposed for a subset of XBs containing unusually massive stellar-mass BHs \citep{deMink+2009}, but any metallicity dependence of the rate for this alternative formation channel is so far unclear. From observational evidence, \citet{Crowther+2010} point out that the BH masses in known wind-accreting BH HMXBs do seem to be larger at lower metallicity.

We are not aware of an observed sample of ULXs with population II metallicity, so the predicted range of ULX metallicity dependence is currently hard to test.  However, \citet{ZampieriRoberts2009} concluded that even in the available local samples of ULXs there are hints of a metallicity dependence \citep[see also][]{Mapelli+2009, Mapelli+2010}. In addition, \citet{ShtykovskiyGilfanov2005} compared the small magellanic cloud (SMC) to the \citet{Grimm+2003} relation between HMXB populations and the SFR, as derived for more massive, higher-metallicity galaxies. Using standard far-IR, H$\alpha$ and UV indicators of the SFR, the HMXB population in the SMC appears be overabundant by a factor of $\sim$10 (see also Dray 2006). If that discrepancy is due to a metallicity dependence, then it is significantly \emph{larger} than the one we apply in this paper.

\begin{figure}
\epsfig{file=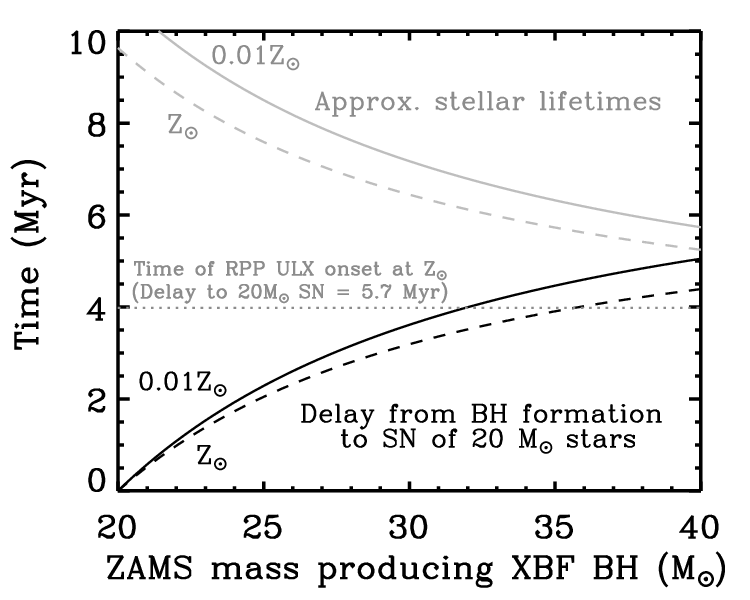, width=8cm}
\caption{ The black curves show estimates of the time available between BH formation and the start of the most energetic SNF, {as a function of the initial stellar mass which produces the BHs, and assuming that SNF arises from stars below $\rm 20 M_{\odot}$ (see text)}. During this delay XBF could operate before SNF. The grey curves show the approximate lifetimes of massive stars up to $40 M_{\odot}$, {from which the black curves are calculated}. The horizontal dotted grey line represents the (shorter) delay between star formation and ULX onset found by RPP, which indicates that we may have underestimated the time gap available for XBF before SNF.
\label{fig:SNdelay} }
\end{figure}

\subsection{The first supernova may not be the first \emph{energetic} supernova: quiet black hole formation}
\label{sec:quietBH}

In a given starburst, BHXBs may be able to provide energy input before the first energetic supernova occurs. From slightly below solar metallicity and lower, the majority of single stars more massive than $\rm \approx 40~M_{\odot}$ are expected to produce black holes by direct collapse, and the associated SNe are expected to produce no significant explosion, perhaps contributing no energy at all to feedback \citep[see, e.g.,][]{Fryer1999, Mirabel+2003, Heger+2003, EldridgeTout2004}.
SNe which produce BHs by fallback may also be very faint and contribute little to supernova feedback \citep[a possible example of such a supernova is 1997D, for which see][]{Turatto+1998}. At a metallicity of $\approx 0.001$, figure 5 of \cite{EldridgeTout2004} indicates that direct collapse occurs for a ZAMS mass of $\rm \gtrapprox 35~M_{\odot}$, and BH formation by fallback for masses $\rm \gtrapprox 20~M_{\odot}$. This also means that the normalisation of the energetic SN rate to the SFR and that of the \emph{total} SN rate to the SFR are not identical.  So most stars $\rm \gtrapprox 20~M_{\odot}$ may hardly contribute to feedback.

Of course, the previous description of events which contribute to SNF is simplified. Firstly, binary evolution makes the situation more complex \citep[e.g.~][]{Podsiadlowski+1992, DeDonderVanbeveren2003, EldridgeThesis, Eldridge+2008}.   In addition, \emph{extremely} massive stars are expected to produce pair-instability supernova, especially at low metallicity \citep{Barkat+1967,RakavyShaviv1967,Heger+2003}, and at least one convincing candidate for such an explosion has been observed \citep{Gal-Yam+2009}.  Recent examples of unexpectedly luminous SNe also indicate that further exceptions to the simple picture above probably exist \citep{Smith+2007,Ofek+2007,Quimby+2007,Smith+2008,Gezari+2009,Miller+2009,Drake+2010,Rest+2011,Quimby+2011}.  These have often been linked with luminous blue variable progenitors, i.e. stars significantly more massive than $\rm 20~M_{\odot}$, although that is not consistent with prior expectations. We note that the leading explanation for the high observed luminosity of most of these events involves a SN which is more efficient at \emph{radiating away} kinetic energy, not an intrinsically more energetic SN \citep{Smith+McCray2007,Agnoletto+2009, Smith+2010}.  Those exceptionally luminous events are rare, although \citet{Gal-Yam+2007} suggest that many more events -- all Type IIn SNe, i.e.\ $\approx 2$ per cent of all bright core-collapse SNe -- might be from such massive stars.  It is possible that a non-negligible fraction of SNF begins sooner after a starburst than suggested by the arguments in the previous paragraph.  

Despite those caveats, we consider it is very plausible that, in the majority of starbursts, the first BHs could be formed well before many `loud' supernovae from that starburst explode and drive significant SNF, especially at low metallicity and hence in the early Universe.  In principle, the first BHXB could switch on almost as soon as the first of those BH is born. The limit is the evolutionary timescale of the donor star: mass transfer should be stable from a radiative donor up to a few times more massive than the mass of the BH. 
Figure \ref{fig:SNdelay} indicates the approximate amount of time which might be available to XBF before strong SNF begins.
Stellar lifetimes are estimated using equations 4 and 57 of \citet{Hurley+2000}. This suggests that if SNe from progenitors more massive than $\rm 20~M_{\odot}$ contribute little to feedback then XBF should have several Myr of free time to act before SNF begins. If the time of ULX onset found by RPP is correct, then there is almost 6 Myr left before $\rm 20~M_{\odot}$ progenitors explode and $\approx$1 Myr before 40 $\rm M_{\odot}$ progenitors do. During this time XBF could modify the environment on which SNF acts. 

Note that, in adopting the work of RPP, we have implicitly adopted their assumption that BHs are born with a mass of $\rm 10~M_{\odot}$, along with their limited range of donor masses; we expect that the time before the first ULX switches on is likely to be even shorter than the minimum time given by RPP.  Moreover, if stellar BH masses increase with decreasing metallicity then the maximum stable donor star masses also increase, which could further shorten the time before XBF can begin.

\begin{table}
\centering
\caption{\label{tab:snfrac} Approximate supernova feedback efficiency necessary for starburst gas ejection. } 
\begin{tabular}{ccc}
\hline 
\hline 
$V_{\rm virial}$ & \multicolumn{2}{c}{ $|U| / \Sigma E_{\rm SN}$} \\
($\rm km~s^{-1}$) & $M_{\rm tot}/M_{\rm stellar}$=5 & $M_{\rm tot}/M_{\rm stellar}$=50 \\
\hline 
100 & 0.1 & 1 \\ 
10 & $10^{-3}$ & 0.01 \\
1 & $10^{-5}$ & $10^{-4}$ \\
\hline 
\hline 
\end{tabular}
\end{table}

\subsection{Gas retention in dwarf galaxies}
\label{sec:gasretention}

Feedback in some dwarf galaxies seems to be able to truncate SF without ejecting all the gas, allowing extended or repeated SF episodes even in some astonishingly tenuous dwarf galaxies \citep[e.g.\ Leo T, for which see][]{Irwin+2007, RyanWeber+2008, DeJong+2008, GrcevichPutman2009}. Many dwarf galaxies do lose their gas \citep[for a review see, e.g.,][]{Tolstoy+2009}, but the evidence that some do not is much more surprising, since for galaxies with virial velocities $\lesssim$10 $\rm km ~s^{-1}$ even very low SNF efficiencies could suffice to unbind the gas. We illustrate this in Table \ref{tab:snfrac}, in which we give order-of magnitude estimates for the efficiency of SNF which would be necessary to eject all the gas from low-mass starburst galaxies. We show estimates for two different ratios of total mass  ($M_{\rm tot}$) to stellar mass  ($M_{\rm stellar}$), one of which represents an extremely dark-matter dominated case. We simply relate the binding energy ($|U|$) to the virial velocity ($V_{\rm virial}$), taking $|U|=M_{tot}~V_{\rm virial}^{2}$. Our estimates also assume that one SN contributes to SNF for every 100 $\rm M_{\odot}$ of stars formed in the starburst, and that perfect SNF efficiency provides $\rm 10^{51} erg$/SN. The important point is that SNF could be less than 1 per cent efficient and would still eject the gas from realistic dwarfs with virial velocity $\rm \lesssim 10~km~s^{-1}$.

Energy injected on the long timescale of XBF input ($\sim$10$^{8}$ yr) seems more likely to be able to heat the gas reservoir non-destructively than energy injected rapidly compared to the dynamical timescale of the galaxy (as in SNF).  Since we also expect that XBF can begin before strong SNF, we furthermore suggest that XBF might create `chimneys' to help channel subsequent SNF away.  Previous studies of how SN explosions eject or enrich gas reservoirs from dwarf galaxies have sometimes found it possible that the hot, metal-enriched gas can escape through such chimneys whilst leaving more of the cold phase intact than simplistically expected (see especially \citealt{MacLowFerrara1999}; also, e.g., \citealt{DeYoungGallagher90}; \citealt{DeYoungHeckman94}; \citealt{Fragile+2003}). Such previous work has found that SN remnants preferentially expand through regions containing hot, low-density gas than through the colder, higher-density phase outside this `chimney', and the SN ejecta can thereby be channelled out of the galaxy.  Instead of a situation where early SNF creates chimneys for later SNe (as in previous work), XBF may be able to form such hot, low-density regions before significant SNF begins.  Even if XBF simply pre-heats the ISM before SNF, without helping to form chimneys, then it might lead to a reduction in SNF efficiency \citep[see, e.g.,][and references therein]{CeverinoKlypin2009}.

Arguments in favour of a relatively gentle feedback process in dwarf galaxies were made by \cite{Lee+2006}, who suggested that `a less energetic form of metal-enhanced mass loss than blowouts could explain the small scatter in the $L-Z$ and $M_{*}-Z$ relationships'; XBF may help fulfil that need. Furthermore, the stochasticity of XBF, as presented in Section \ref{sec:fluctuations}, may help explain the diversity in the known Milky Way satelite galaxies, e.g. how Leo T can differ so much from the others in present-day neutral gas content.

\subsection{Timescales \& episodic star formation}
\label{sec:episodic}

A mechanism which stops star formation (SF) without ejecting the gas reservoir may allow episodic SF.  To explain separated bursts of SF  seen in some dwarf galaxies, \citet{QuillenBlandHawthorn2008} suggest that the feedback in those galaxies should be delayed over a timescale of 50--100 Myr. The delay timescale is not the only important factor in their model, but they find that such a delay is necessary to produce episodic star formation.  As there seemed to be no stellar source of feedback with a suitable timescale, they concluded that the explanation for this timescale must be non-stellar. Based on the models of RPP, energy input from XBF can increase for $\sim$10$^{8}$ years after the onset of SF, apparently providing a stellar source of feedback naturally consistent with that model for episodic SF (see also Section \ref{sec:XBenergy}).  Continuing energy input from longer-lived XBs might further delay the subsequent star formation episodes.  

Stellar feedback could perhaps be delayed in another way such that it could satisfy the conditions found by \citet{QuillenBlandHawthorn2008}; they themselves speculate on how the effective timescale of SNF might be extended. Some of the feedback from SN Ia could happen on a sufficiently long timescale and some core-collapse SNe could be delayed by the effect of binary evolution \citep[see, e.g.,][]{DeDonderVanbeveren2003,Eldridge+2011}. 
The model of \citet{QuillenBlandHawthorn2008} is not the only attempt to explain how episodic SF can occur. For example, the simulations of \citet{PontzenGovernato2011} produce bursty star formation in a dwarf galaxy without modifying their input from stellar feedback. Nonetheless, based on the \citet{QuillenBlandHawthorn2008} model, XBF may help to explain this effect; we also repeat that it seems hard for simple SNF to allow gas retention in galaxies such as Leo T.

The extended duration of energy input from XBF also means that XBs born with low metallicity in a galaxy with a large specific SF rate could still be active when the star-forming metallicity of the galaxy has already been enriched; this could be important if the XBF input is strongly metallicity dependent.

\begin{figure*}
\begin{centering}
\epsfig{file=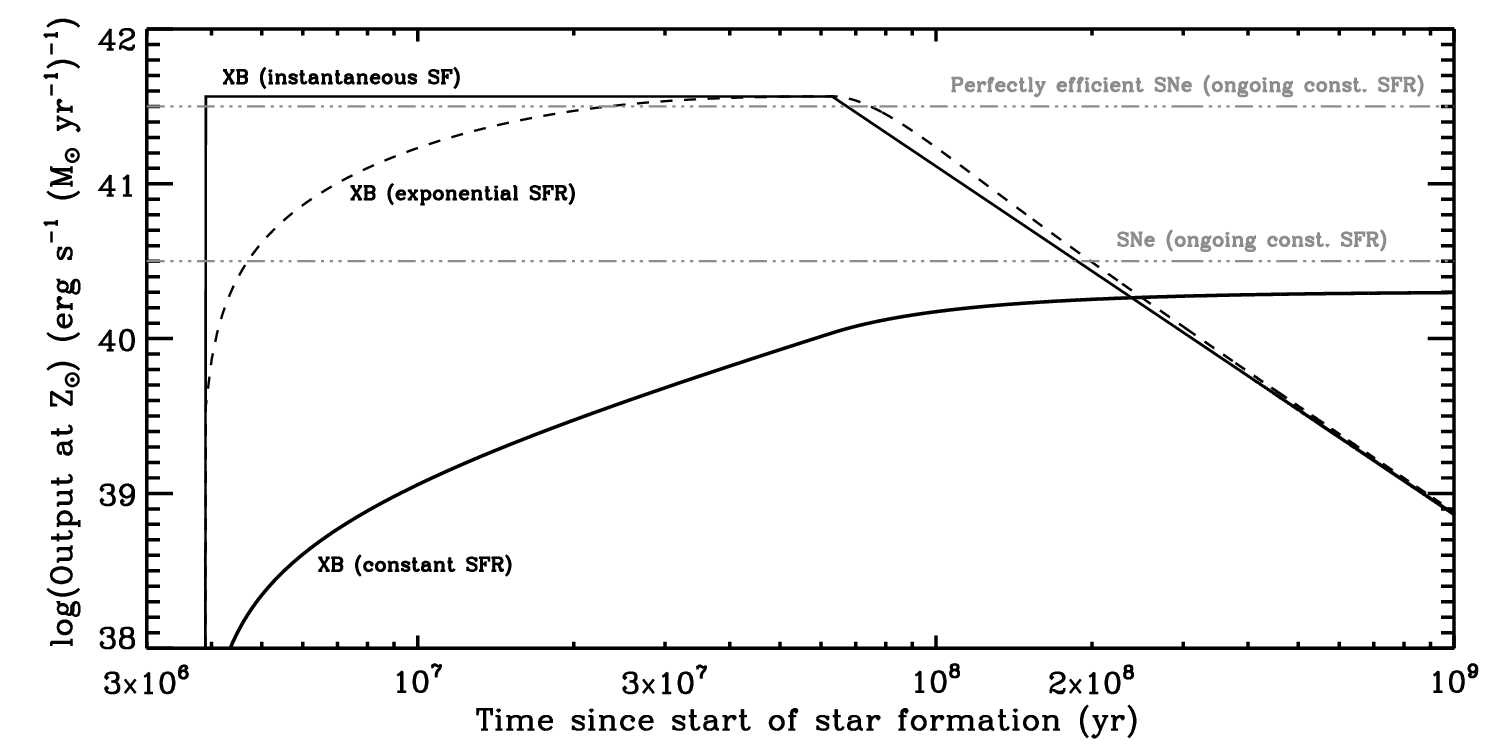, width=0.9\textwidth}
\caption{\label{fig:powercomparison} The three black curves show the energy output from the young XB population of three different SFHs, i.e. an instantaneous burst of SF, a constant SFR, and an exponentially decaying SFR. Each simple population contains a total SF mass of $\rm 10^{9} M_{\odot}$, at constant metallicity ($Z_{\odot}$).  For comparison, we also show grey lines showing the SN energy input (assuming one SN per 100 $\rm M_{\odot}$ of SF, and $\rm 10^{51} erg$ of energy per SN), but for `ongoing', i.e. pre-existing, constant SF rather than for the new SF which produced these XB populations; for a discussion of the likely delay to SNF, see section \ref{sec:quietBH} and Figure \ref{fig:SNdelay}. The standard SNF line assumes the canonical 10 per cent efficiency.}
\end{centering}
\end{figure*}

\section{Mean X-ray binary populations}
\label{sec:XBenergy}

Here we present the expected energy output from mean X-ray binary populations, as measured by their X-ray luminosities.  We have already argued that the efficiency with which these outputs might couple to the ISM could be very high (see Section \ref{sec:efficency}).  The potential importance for galaxy evolution of the mean energy inputs described here are illustrated by scaling relations in the appendix. However, much of the later sections discuss how these simple treatments alone may not capture the essence of XBF, especially for low mass or low star formation rate (SFR) populations.  

\subsection{X-ray binaries in young populations}

\cite{Grimm+2003} observationally found that the most luminous sources dominate the X-ray output of a star-forming galaxy, so here we can concentrate on those most luminous sources when describing the power output from young XB populations. We use the theoretical models of RPP, which assume that those dominant X-ray sources are ULXs formed from stellar-mass BHs, but we also demonstrate that those are consistent with observations.  In Section \ref{sec:fluctuations} we  consider the situation at low absolute SFR, when it is unlikely that the most luminous source in a galaxy is a ULX.

Scaled to a steady-state SFR of 1 core-collapse SN per century, RPP's models produce a total accretion luminosity from ULXs, ($L_{\rm ULXs}$) of $2.3 \times 10^{41} {\rm erg~s^{-1}}$ if the Eddington luminosity can be exceeded by a factor of 10, and $1.5 \times 10^{41} {\rm erg~s^{-1}}$ if the Eddington luminosity is strictly obeyed. Their calculation was performed at solar metallicity ($Z_{\odot}$), and assumed BHs of mass $\rm 10~M_{\odot}$.  Whilst these calculations are subject to several uncertainties, RPP seemed to reproduce well the observed ULX distribution in the Cartwheel galaxy. Based on their results at $Z_{\odot}$ we assume:
\begin{equation}
\label{eq:RPP05}
L_{\rm XBF} \approx 2 \times 10^{40} {\rm~erg~s^{-1}}  \left(\frac{SFR}{\rm M_{\odot}~yr^{-1}}\right)
\end{equation}
for the luminosity of the entire XB population in a mean star-forming galaxy. 
This seems consistent with the relationship between SFR and total X-ray luminosity in the 2-10 keV band
($L_{\rm X,2-10keV}$) found by \cite{Grimm+2003}. In the linear regime (i.e.\ for mean populations), their empirical relation is:
\begin{equation}
L_{\rm X,2-10keV} = 6.7 \times 10^{39} {\rm~erg~s^{-1}} \left(\frac{SFR}{\rm M_{\odot}~yr^{-1}}\right) 
\end{equation}
which would be equivalent to equation \ref{eq:RPP05} if $\approx 2/3$ of the bolometric luminosity is not observed in the 2--10 keV band. It also seems likely that the SFR normalisations are not identical when comparing the empirical to the theoretical luminosity.

We take Equation \ref{eq:RPP05} as potentially giving the energy input to XBF at $Z_{\odot}$. This is somewhat less than the approximate energy output of one SN ($\approx 10^{51}$ erg) per century ($\approx 3 \times 10^{9}$ s), i.e.\ $\approx 3 \times 10^{41} {\rm~erg~s^{-1}}$, which is roughly appropriate for a SFR of $\rm 1~M_{\odot}~yr^{-1}$ (depending on the IMF).  However, XBF would become competitive if the efficiency of XBF is greater than of SNF (as previously suggested).  Furthermore, since we expect that the importance of ULXs increases as the metallicity decreases (see Section \ref{sec:metallicityintro}) then ULXs could easily dominate SNe as an energy source for feedback as the metallicity decreases.

\subsection{X-ray binaries in old populations}
\label{sec:oldpops}

LMXBs may also provide useful energy input in old populations to keep gas hot. The \cite{Gilfanov2004LMXB} relation:
\begin{equation}
\label{eq:lmxbs}
  L_{LMXB} \approx 10^{37}~{\rm erg~s^{-1}~}\left( \frac{ M_{*}}{10^{8} ~{\rm M_{\odot}}} \right)
\end{equation}
implies significant energy input in a massive elliptical (see the scaling relations in the appendix for estimates of how much gas might be kept hot by XBF in old populations).

The metallicity dependence of this energy input is less clear than for ULXs, though there is no expectation that it should be as strong as for young BHXBs. For globular clusters (GCs), XB luminosity is higher in metal-\textit{rich} GCs than metal-poor ones, but the origin of this effect is unsolved \citep[see, e.g.,][]{Bellazzini+1995, Kundu+2002,  Jordan+2004, Ivanova2006, Kundu+2007, Sivakoff+2007}.

\subsection{Mean X-ray binary feedback prescriptions for semi-analytic models}

We now suggest a simple treatment of XBF intended for use in semi-analytic models, based on our earlier descriptions. Here we assume that XBF efficiency (measured with respect to XB luminosity; see Section \ref{sec:efficency}) is 100 per cent, but clearly another parameter could be added.  

The time- and metallicity-dependence of the energy input from young XBs makes their inclusion slightly complicated.  We have produced a fit to the energy output from a ULX population produced by an instantaneous burst of star formation (using model B in figure 8 of RPP).  This fit is in two parts. We use the fitting function $L_{\rm XBF,1}$ for the XBF energy input at early times, when there is a high-luminosity plateau in the power output.  At later times, we use $L_{\rm XBF,2}$ to represent the decline of the XB power output as the population ages. For even older stellar populations, longer-lived XBs should set a floor on the XBF energy input.  For mean populations, Equation \ref{eq:lmxbs} could be used to give the minimum energy input from XBs at any age. 

For any star formation history (SFH) other than an instantaneous burst then at some time $t_{\rm now}$ the instantaneous power input from XBF will be a combination of contributions from populations with different ages.  For young populations ($\rm \lesssim 1~Gyr$) the XB power output should be the sum of both $L_{\rm XBF,1}$ and $L_{\rm XBF,2}$, where:
\begin{equation}
\label{eq:PowerInputOne}
\frac{L_{\rm XBF,1}}{\rm erg~s^{-1}} \approx 2 \times 10^{40} \int_{t_{\rm now}-t_{1}}^{t_{\rm now}-t_{2}}
  \frac{SFR(t')}{\rm 1~M_{\odot}~yr^{-1}} 
  \times f( Z(t') ) dt'
\end{equation}
and $SFR(t')$ {and $Z(t')$} were the star formation rate and {metallicity} \emph{of the newly-forming stars}
{at} some previous time $t'$.
We require $f({\rm Z}_{\odot})=1.0$, and suggest setting
$f({\rm Z}_{\odot}/200)=10$.  Note, however, that this function is uncertain; further observational and theoretical work is necessary to determine it more accurately. 

We estimate that $t_{1}=10^{6.6}$ and $t_{2}=10^{7.8}$ years (again using model B in figure 8 of RPP). As explained in Section \ref{sec:quietBH}, this value for $t_{1}$ could probably be reduced. At later times we approximate the power input from this young component as steadily declining:
\begin{equation}
\label{eq:PowerInputTwo}
\frac{L_{\rm XBF,2}}{\rm erg~s^{-1}} \approx  10^{57.85} \int_{t_{\rm now}-t_{2}}^{t_{\rm now}-\infty} {(t_{\rm now}-t')}^{-2.25} \frac{SFR(t')}{\rm 1~M_{\odot}~yr^{-1}} dt'
\end{equation}
(we have suppressed the metallicity dependence here for brevity, but presume that the same function as used in Equation \ref{eq:PowerInputOne} could be applied here).

Figure \ref{fig:powercomparison} shows the outcome of the above equations at solar metallicity.  We re-emphasize that, at low metallicity, the relative importance of XBF in young populations should increase. As a result then, if the self-enrichment time of a particular galaxy is shorter than $\sim 10^{8}$ years, then the \emph{total} feedback from XBF might easily still be \emph{increasing} even when the mean metallicity of the stars in the galaxy is no longer low.

\begin{figure}
\epsfig{file=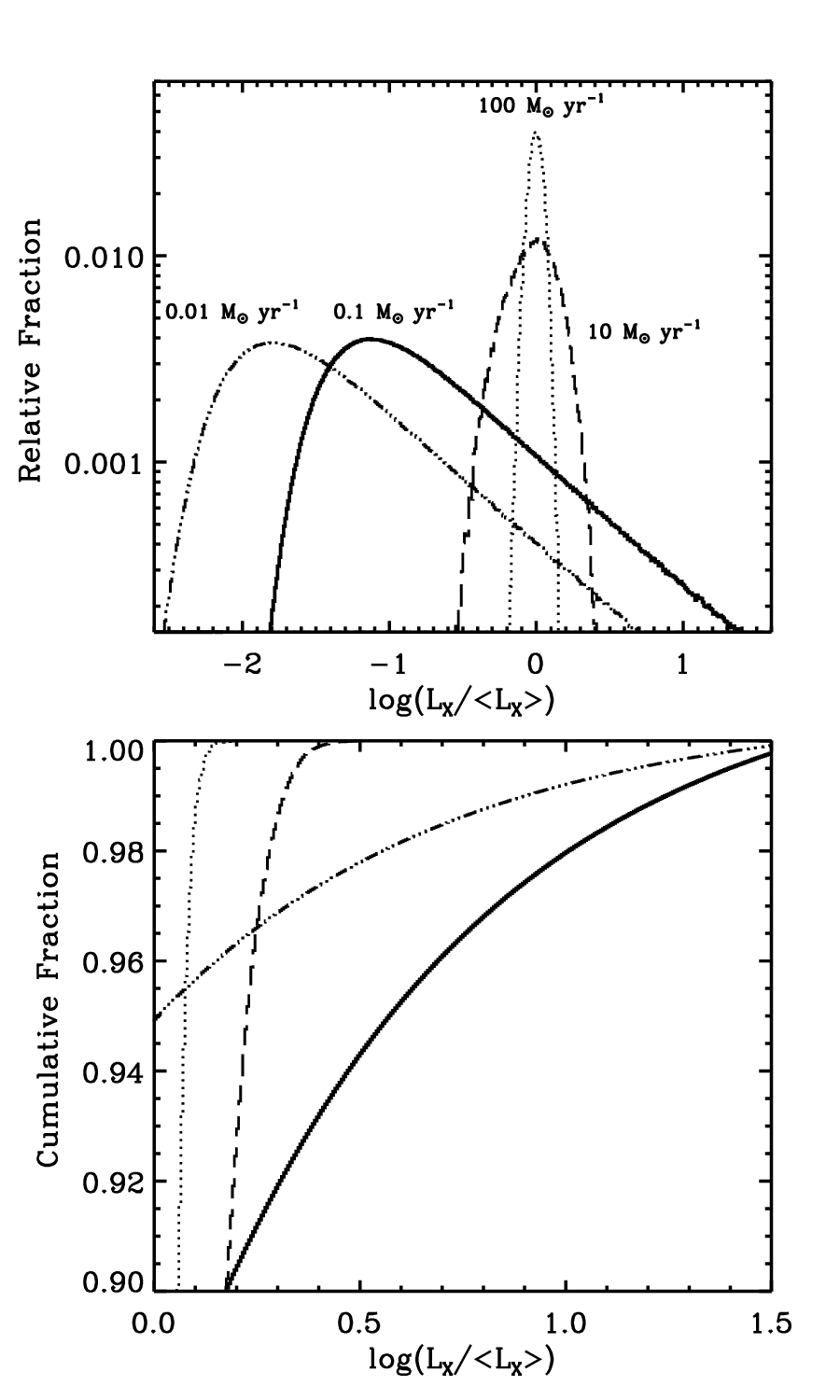, width=8cm}
\caption{ The top panel shows how the statistical variation in XBF populations may be realised for different absolute SFRs (at $\rm Z_{\odot}$); see text for details. The lower panel uses the cumulative distribution of the same population models to show the increased importance of the \textit{high}-luminosity tail of the distribution for \textit{low} SFRs. As the mode of the distribution falls for low SFR, the effective range of XBF input in low-SFR halos extends over two orders of magnitude, even ignoring metallicity effects.
\label{fig:statistics}
} 
\end{figure}

\section{The importance of stochasticity}
\label{sec:fluctuations}

Galaxies with low absolute mass and SFR are likely to deviate from the mean relations given in the previous section. The effect that random sampling of the initial mass function has on observed stellar populations has been investigated by many authors \citep[see, e.g.,][]{Cervino+2002, Cervino+2003, Villaverde+2010, daSilva+2011, EldridgeIMF2011}.  When considering the formation of XBs, the parameter space of \emph{binary} initial conditions is multi-dimensional, and the volume of that space which produces bright XBs is much smaller than the volume which leads to a massive star and a supernova.  So we expect stochastic variation to be greater for XBF than for SNF, especially because the power output from XBs should be dominated by a small fraction of the most luminous sources.  Here we estimate the potential importance of this effect, in particular by comparing the distribution of $L_{\rm X}$, the XBF luminosities for each population within an ensemble, to $<$$L_{X}$$>$, the global mean value.

At the peak luminosity of the population synthesised by RPP, the number of luminous BHXB systems in a starburst which produces $10^{6}$ SNe is  $\approx$30, whilst at an age of $\rm 10^{7}~yr$ that population only contains $\approx$10 sources with $L\gtrsim 10^{39} {\rm erg~s^{-1}}$ (there is a small model-to-model variation between the numbers of expected systems, as shown in their figures 6 \& 7, but not one which is large enough to affect this argument).  \emph{Because the majority of the expected XBF luminosity at those early times is produced by far fewer sources than lead to the SNF energy input, statistical variation between populations at these times should be more significant for XBF than for SNF.}

\citet{Gilfanov+2004statistics} showed more formally that the variation between young XB populations increases significantly as the SFR decreases. 
They analytically combined the probability distribution function for the number of sources present with the luminosity function from which those sources are drawn; they thus found the probability distribution for the total luminosity (below we demonstrate this numerically). They found that the regime in which small-number statistics affects the expected total luminosity extends much further than might naively have been expected; e.g.\ for populations of young X-ray binaries it can extend to SFRs of a few solar masses per year.  This may well have implications for the evolution of populations of low-mass galaxies, in particular since the probability distribution functions derived by Gilfanov et al.\ become strongly asymmetric at low SFR.

\subsection{Low star-formation rate effects for young populations}
\label{sec:ULXflucts}

For the reasons given above, it seems possible that stochastic differences in XBF magnitude between galaxies with similar initial properties -- i.e.\ random halo-to-halo variations -- might be important when trying to understand the effects of XBF.  Here we estimate the potential significance of XBF stochasticity in young populations by producing a probability distribution function for the XBF magnitude.  To do this we use a Monte-Carlo realisation of the work of \citet{Gilfanov+2004statistics}, i.e.\ we populate each simulated starburst with a collection of XBs which are randomly selected from an X-ray luminosity function (XLF) in such a way as to preserve the appropriate mean for the total luminosity in the large-SFR limit.  We adopt their power-law XLF with an index of $-1.6$, and assume that no ULX can be more luminous than $\rm 10^{41}~erg~s^{-1}$. We further allow the XLF of the individual sources to extend over eight orders of magnitude for these young populations; this choice affects the details of the results, but not the general principle. Specifically, reducing this range over which the power-law operates increases the fraction of galaxies with no XBF whatsoever, rather than with little XBF; that makes for uglier plots, but does not change the qualitative conclusions.  Figure \ref{fig:statistics} shows the outcome of those calculations, demonstrating that the bulk of halos with low absolute SFR could put out X-ray luminosities an order of magnitude below the mean value, $<$$L_{\rm X}$$>$, whilst several percent of initially similar halos could experience XBF an order of magnitude stronger than the mean value. This two-order-of-magnitude spread is larger than the systematic increase in mean XBF magnitude, by a factor of roughly 10, which we expect to occur at low metallicity.  

For low-mass, low-absolute-SFR galaxies at low metallicity, the effects of metallicity and stochasticity should combine, meaning that for a small percentage of halos the \emph{specific} XBF (relative to the SFR) could be \emph{two} orders of magnitude higher than the mean value given by Equation \ref{eq:RPP05}.  Note that as the expected number of bright sources increases within the galaxy as the metallicity drops then the spread at a given SFR also decreases; ignoring any contribution from higher mean BH masses then the shape of the $L_{\rm X}$ distributions is constant for total $<$$L_{\rm X}$$>$. It is possible that the shape of the XB luminosity function is also a function of metallicity; if so this may be as important as the metallicity dependence of the XBF normalisation.

In high-absolute-SFR populations, the likelihood of a large deviation from the mean is small, and the mean relations from section \ref{sec:XBenergy} can be applied. The importance of stochasticity to XBF should also be somewhat reduced for situations in which the significant input from XBF can be averaged over a relatively long time, i.e.\ over several generations of luminous XBs.  However, when trying to understand the diversity in the baryonic content of low-mass halos, we suggest that probability distributions for XBF, such as that described above, should be adopted. We emphasize that this should be true even for high \emph{specific} SFRs.

\subsection{Low mass effects for old populations}
\label{sec:LMXBflucts}

We do not repeat the above calculations for LMXBs, as the remainder of this paper will be mostly concerned with XBF in young populations. However, we will briefly consider when XBF in old populations might be affected by significant stochastisicity.  Assuming a quantised LMXB luminosity of $\rm 10^{35}~erg~s^{-1}$,  Equation \ref{eq:lmxbs} gives a rough expectation value of one LMXB for a mass of $\rm 10^{6} M_{\odot}$. The more sophisticated analysis of \cite{Gilfanov2004} finds an expected number of LMXBs with luminosity $> \rm 10^{37}~erg~s^{-1}$ of $\approx$ 140 per $\rm 10^{11} M_{\odot}$. Based on \cite{Gilfanov2004}, we suggest that using mean values of LMXB XBF might be misleading for galaxy masses much less than $\rm 10^{11} M_{\odot}$, so a PDF for `old' XBF may also be needed to faithfully reproduce systems where LMXBs are significant.  

As canonical LMXB lifetimes are $\approx$1 Gyr, then any dwarf galaxies low enough in mass to posess (typically) one significant LMXB could be expected to experience heating episodes lasting from a few hundred Myr to several Gyr. Cooling and a new epoch of star formation might be delayed until the LMXB heating stops.  Scaling relations in the appendix suggest that it may be reasonable for a baryonic gas fraction of several tens of percent to be kept hot in some low-mass galaxies by XBF.  The inherent randomness of XBF may again be important in understanding the present-day diversity of dwarf galaxies, and should be considered in future models.

\begin{figure}
\epsfig{file=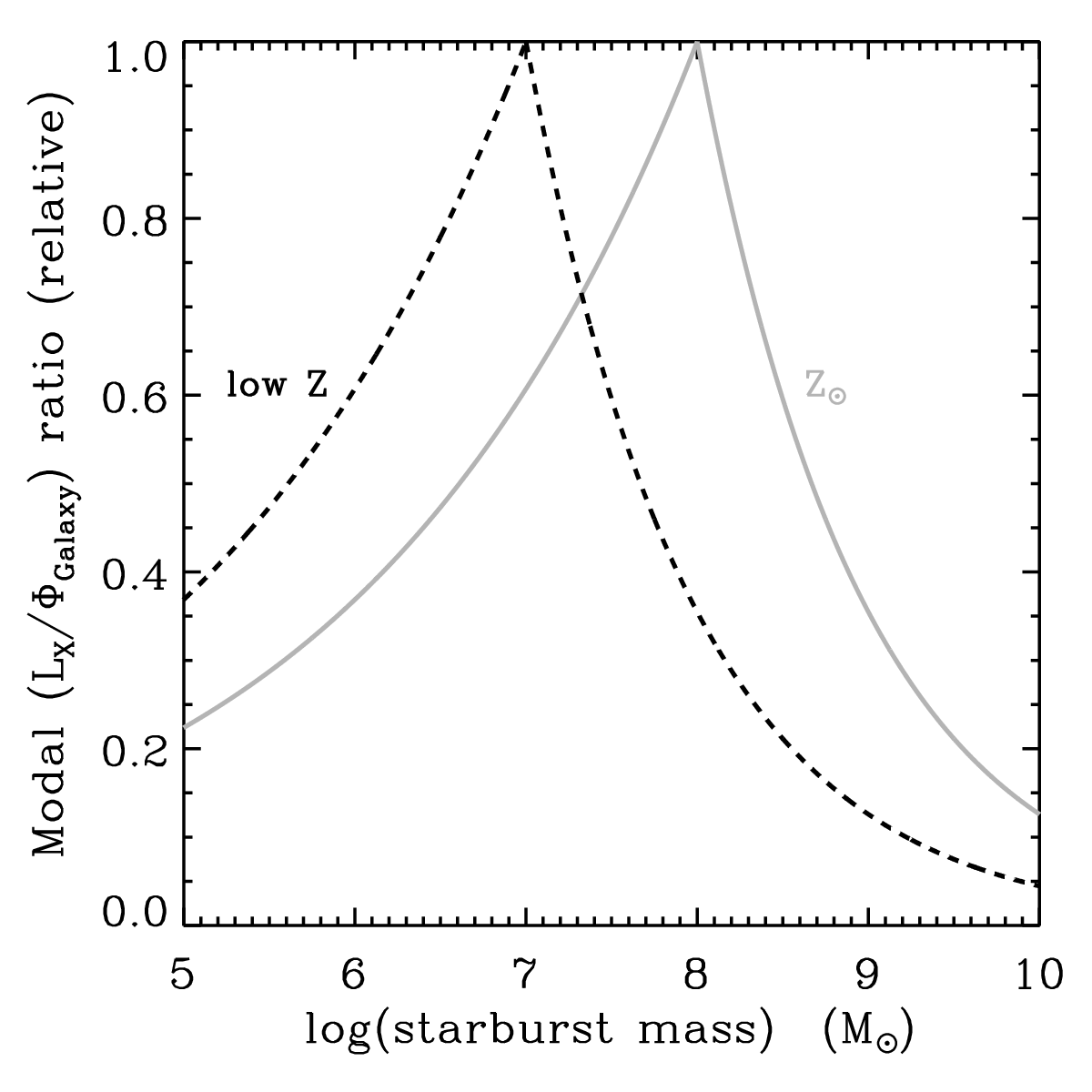, width=7cm}
\label{fig:commonsignif}
\caption{ We estimate the relative importance of XBF for modal galaxies in the XBF probability distribution by taking the ratio of the modal XBF luminosity ($L_{\rm X}$) to the gravitational binding energy ($\rm \Phi_{\rm Galaxy}$) for each galaxy mass. We expect that the low-Z curve (broken black line) should be relevant for the galaxy mass range at which this ratio peaks. As this figure only indicates the significance of XBF for the mode of the galaxy population, we repeat that the shape of the XBF probability distribution itself is relatively narrow and roughly symmetric for galaxies more massive than those at the peak in this figure, but increasingly broad and asymmetric for less massive galaxies (see Section \ref{sec:ULXflucts} and Figure \ref{fig:statistics}).} 
\end{figure}

\section{For which galaxy masses should X-ray binary feedback be most commonly significant?}
\label{sec:peakdominance}

As explained in the previous section, then below a certain absolute SFR the \emph{most probable} XBF input for a galaxy with a given SFR should be below the expectation value for XBF energy input at that SFR. That is, the mode of the XBF probability distribution is below the mean.  In that low-SFR regime, the location of the mode scales as $\rm (SFR)^{5/3}$ for an XLF slope of $-1.6$ (see Section \ref{sec:ULXflucts} and \citealt{Gilfanov+2004statistics}).  Above this regime, i.e.\ for high enough SFRs that the mode of the $L_{\rm X}/$ distribution is very close to $<$$L_{\rm X}$$>$, then clearly the most likely XBF input scales linearly with the SFR.  This allows us to make a simple estimate of which part of the galaxy population might be the most likely to be modified by XBF.

We will assume that the gravitational binding energy typically scales as $M^{1.45}$ \citep[e.g.][]{Saito1979, YoshiiArimoto1987}, and that we can take a roughly constant star formation lifetime $\Delta t_{\rm SF}$. In this case then the galaxy mass where the ratio of XBF energy input over gravitational binding energy is highest -- for a typical halo, rather than for the tail of the distribution -- will arise at the transition between these low-SFR and linear regimes.  We can use these assumptions to estimate the ratio between the XBF luminosity and galaxy binding energy for `typical' galaxies of different starburst masses. Figure \ref{fig:commonsignif} shows these estimates, assuming a single dominant starburst with a duration of $10^{8}$ years.

Using our previous assumptions, and writing the effective upper end of the XLF as $L_{\rm max}$ then the stellar mass $M_{*}$ corresponding to the point of most common, most significant XBF is estimated to be:
\begin{equation}
\left( \frac{M_{*}}{\rm 5 \times 10^{7} M_{\odot}} \right) \sim \left( \frac{L_{\rm max}}{1 \times 10^{40} \rm erg~s^{-1}} \right) \left( \frac{\Delta~t_{\rm SF}}{4 \times 10^{7} \rm yr} \right)
\end{equation}
at solar metallicity, or:
\begin{equation}
\left( \frac{M_{*}}{\rm 5 \times 10^{6} M_{\odot}} \right) \sim \left( \frac{L_{\rm max}}{1 \times 10^{40} \rm erg~s^{-1}} \right) \left( \frac{\Delta~t_{\rm SF}}{4 \times 10^{7} \rm yr} \right)
\end{equation}
if we allow a factor of ten enhancement to the number of bright XBF due to low metallicity. (The lower mass scale of Equation 7 could also be appropriate at solar metallicity if, in Section \ref{sec:fluctuations}, we have overestimated the significance of stochasticity on XBF by an order of magnitude for any given SFR.)
Such stellar masses are consistent with the masses identified by \cite{Sawala+2011} for which feedback recipes in current galaxy models may be inadequate. More sophisticated investigation of this seems worthwhile. We stress that here we have estimated the mass scale at which significant XBF seems likely to be common in the greatest fraction of galaxies with that given mass. This is distinct from the content of Section \ref{sec:ULXflucts}, in which we argue that a few percent of the lowest-mass halos should experience the greatest specific impact of XBF.

\begin{figure*}
\epsfig{file=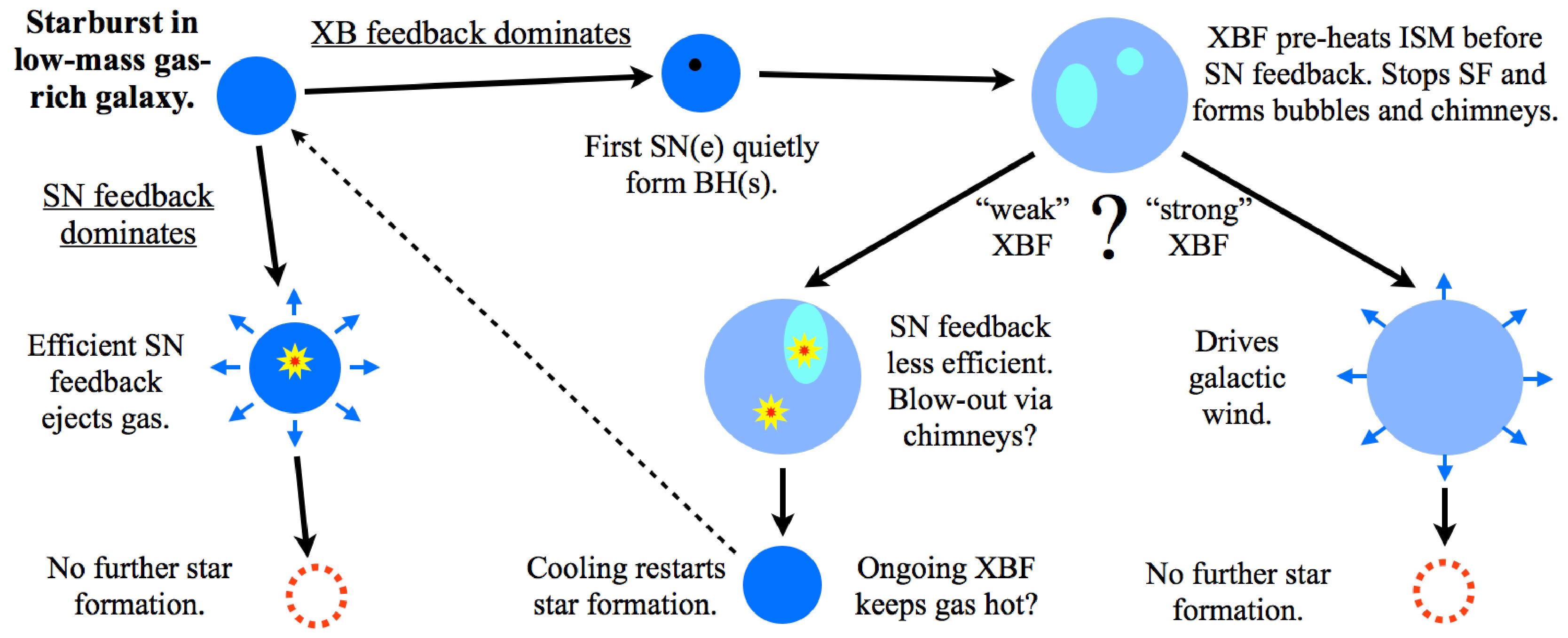, width=18cm}
\caption{ Schematic showing speculations on the potential interplay between XB and SN feedback.
\label{fig:schematicinterplay}
} \vspace{3ex}
\end{figure*}

\section{Discussion}

Feedback from XBs has properties that should not be ignored when modelling the formation and evolution of some galaxies.
In particular, the long, slow energy input seems ideal to explain gas retention and episodic star formation in some low-mass dwarf galaxies. Even though metallicity is an important factor, the XBF probability distributions and the interaction of XBF and SNF are potentially even more important. The earliest population of galaxies, initially of very low metallicity \textit{and} low absolute SFR, seem very likely to have been affected by ongoing XBF.

\subsection{The interplay between X-ray binary and supernova feedback}
\label{sec:interplay}

Combining XBF and SNF suggests a rich variety of possibilities, especially given the broad probability distributions described in Section \ref{sec:fluctuations}. We have explained that a few percent of halos with low absolute SFR might be expected to experience a specific contribution from early XBF that is two orders of magnitude greater than the modal XBF input.
Since the observed satellite galaxies may originate in only a similarly small percentage of haloes with low masses it seems reasonable to speculate on whether XBF could help some low-mass galaxies retain gas despite the effect of SNF.   XBF should also apply to the metallicity-dependent disruption probabilities of globular clusters, which we intend to study in a future paper.  XBF-driven expansion could at least decrease the amount of gas within the virial radius at the point when SNe eject the bulk of the gas.

Figure \ref{fig:schematicinterplay} illustrates three scenarios, all of which might occur in initially similar galaxies due to the stochastic variation described in Section \ref{sec:fluctuations}. Two of the scenarios occur when one of SNF or XBF is dominant; in either case the feedback might be strong enough to eject the gas reservoir from the galaxy. The intermediate case, when neither XBF or SNF are dominant, is potentially the most interesting.  If XBF heating is enough to stop star formation, but not to eject the gas, then the conditions into which SN explode are changed. The XBF-created hot bubble(s) seem(s) likely to aid the blow out of the SN ejecta \citep[see, e.g.,][]{DeYoungHeckman94}, and reduce the effective coupling of the SNF to the remainder of the gas reservoir.  

The balance between these three possibilities, and the quantitative conditions which define them, deserve further detailed work.  For example, the development of chimneys and blow-out could be further aided if the distribution of the hot phase is strongly nonspherical, as might be expected for microquasars. Another intriguing detail which should be pursued is the difference in the ratio of energy and momentum involved in SNF and XBF. Moreover, in SS433 the dominant outflows of energy and momentum are perpendicular to each other \citep{Blundell+2001}.

\subsection{A scale length for supernova feedback?}

The effect that XBF pre-heating might have on SNF should be considered in all galaxies. Aware that pre-heating (by massive stars) might reduce the effectiveness of SNF, \citet{CeverinoKlypin2009} suggested that those SNe produced by runaway stars which explode in the colder gas outside star-forming regions could dominate SNF.  However, we note that this sets a natural scale-length. {In some galaxies then stars may not be able to travel far enough to produce effective SNF before leaving the galaxy. }

\citet{Eldridge+2011} modelled populations of runaway supernovae and found that the most common class of supernova, Type IIP, ran away \emph{on average} $\approx$50 pc from their birthplace before exploding (48 pc at solar metallicity, 54 pc at a metallicity of 0.004).  If \citet{CeverinoKlypin2009} are correct and the stars need to runaway to some extent in order to produce efficient SNF, then the average runaway distance might be significantly greater: considering only the secondary stars which produce Type IIP supernovae, i.e. selecting against those which don't run away, \citet{Eldridge+2011} give a mean distance travelled of 190 pc at solar metallicity (and 280 pc at a metallicity of 0.004).  The assumptions used by \cite{CeverinoKlypin2009} led to typical runaway distances between 10 and 100 pc. (A runaway speed of $\rm 20~km~s^{-1}$ for 10 Myr gives a distance of $\approx$200 pc.) Runaway-star SNe might dominate SNF in more massive galaxies, yet decrease SNF in the lowest-mass dwarfs since more SNe take place \emph{outside} the galaxy. 

A similar possibility emerges from a different route: when the size of the hot bubble created by XBF dominates a galaxy, then again SNe seem less likely to be able to have the impact that they otherwise could.  We note, as mentioned above, that the size of the XB-inflated nebula observed by \citet{Pakull+2010} is $\rm \approx 300~\rm pc$, i.e.\ of comparable size to the lowest-mass Milky Way satellite galaxies \citep[see, e.g.,][]{Tolstoy+2009}.

\section[]{Summary and Conclusions}

Stellar physics should not be oversimplified if we desire accurate galaxy models; the physics of the baryonic component matters. We have explained why XB feedback could be important, with characteristics perhaps significantly different from SN feedback.  XBF may well dominate some systems \emph{before} SNF acts, changing the properties of the galaxies into which SNF is later unleashed. XBF from young, luminous systems can still be important $\sim 10^{8}$ years after the formation of the relevant population, by which time the instantaneous star formation metallicity could have greatly increased.  XBF from old populations might also be significant for keeping gas hot in some non-star-forming galaxies.

The properties of XBF might stop star formation but still allow gas retention and multiple star-formation episodes in some very low-mass dwarf galaxies.  The fraction of runaway massive stars might reduce the effectiveness of SNF for galaxies smaller than $\sim$100 pc, even though those runaway SNe might increase the effect of SNF in massive galaxies.  Furthermore, natural statistical variation in the balance between energy input from XBs and SNe could help explain how most low-mass halos eject their gas but some satellite galaxies can retain gas during multiple SF episodes.  The impact of XBF seems likely to be greatest in early-Universe proto-galaxies, when low-mass halos are forming stars with low metallicity. In this case then episodic star formation might be expected in the early epochs of galaxy formation. 

Interplay between different forms of stellar feedback might increase the richness of galaxy evolution, especially for low-mass galaxies where the greatest statistical variation in XBF is expected.  The statistics of XBF in young populations also suggest that we should distinguish between the galaxy masses where the specific XBF should be largest in some individual cases (a tail of low-mass halos with low star formation rate and metallicity) and the galaxy masses where XBF seems most likely to impact a randomly-chosen halo (those which form $\rm \sim 10^{7}~M_{\odot}$ of stars) . 

Probability distributions for metallicity-dependent XBF should be included in future galaxy formation models, especially when studying dwarf galaxies and the earliest epoch of galaxy formation.

\section*{Note Added In Proof}
We are grateful to have been very kindly reminded about the work of \citet{Power+2009}. That paper studied the potential contribution of some young XBs to reionization and therefore also has relevance to early galaxy evolution. 
\emph{(In this version for the arXiv, we have also added footnote 1, which notes \citealt{Power+2009} in Sections \ref{sec:intro} and \ref{sec:efficency}.)}

\section*{Acknowledgements}
 
The authors thank an anonymous referee for generously giving their time and for many very helpful comments; 
in particular, these helped to significantly improve the clarity of our presentation. 
The authors are grateful to numerous people for helpful discussions over several years, in particular
to Philipp Podsiadlowski and Chris Wolf.
SJ thanks many members of and visitors to KIAA for relevant conversations, especially 
Matthias Gritschneder, Eric Peng \& Peter Berczik. SJ has been partially supported by China National 
Postdoc Fund grant number 20090450005 and the 
National Science Foundation of China grant numbers 10950110322 and 10903001. 
Support for the work of KS was provided by NASA through Einstein/Chandra Postdoctoral 
Fellowship grant numbers PF9-00069 issued by the Chandra X-ray Observatory Center, which 
is operated by the Smithsonian Astrophysical Observatory for and on behalf of NASA 
under contract NAS8-03060.


{\small

}

\appendix

\section{Heating and cooling scaling relations}

Here we provide approximate relations for the luminosity necessary to offset gas cooling, in order to illustrate the potential impact of XBF. Nonetheless, we wish to emphasize that kinetic outflows from some XBs suggest that the dominant effect of XBF might not be thermal. Furthermore, averaging XBF input over entire galaxies may also be misleading: the way that XBF locally restructures the ISM before SNF might be its most significant direct consequence (depending on the size of the galaxy in question). 

In what follows we use the mean expected XBF energy inputs at solar metallicity, as presented in Section \ref{sec:XBenergy}, i.e. Equations \ref{eq:RPP05} and \ref{eq:lmxbs} (for the input to young and old populations, respectively).  

Taking the simplifying assumption of a uniform distribution of gas with mass $M_{\rm g}$ in a sphere of radius $r$, we can scale the the cooling luminosity $L_{\rm cool}$ as approximately:
\begin{equation}
\label{eq:cooling}
\frac{L_{\rm cool}}{\rm 10^{36}~erg~s^{-1}} \approx \Lambda_{-22} \left( \frac{M_{\rm g}}{\rm M_{\odot}} \right)^{2} \left( \frac{r}{\rm pc} \right)^{-3} 
\end{equation}
where $\Lambda_{-22}$ is the cooling constant in units of $10^{-22} {\rm ~erg~cm^{3}~s^{-1}}$; this represents very efficient cooling.

\subsection{Keeping gas hot in old galaxies}
If we write the gas mass as a fraction of the stellar mass ($M_{*}$), such that $M_{\rm g} = gM_{*}$, then we can write the following for the ratio of the cooling to the XBF heating for old populations:
\begin{equation}
\frac{L_{\rm cool}}{L_{\rm LMXB}} \approx 10^{7} g^{2} \Lambda_{-22} \frac{M_{*}}{\rm M_{\odot}} \left( \frac{r}{\rm pc} \right)^{-3}
\end{equation}
where $L_{\rm LMXB}$ has been taken from Eq. \ref{eq:lmxbs}.
If we further use the mass-radius relation derived for bright ellipticals in equation 12 of \cite{Hacsegan+2005}, we find:
\begin{equation}
\frac{L_{\rm cool}}{L_{\rm LMXB}} \approx 10^{6.3} g^{2} \Lambda_{-22} \left(  \frac{M_{*}}{\rm 10^{12}~M_{\odot}} \right)^{-0.845}
\end{equation}
so a gas fraction of $g \approx \sqrt{10^{-6.3}} \approx$ 0.1 per cent  might be kept hot in such old massive ellipticals by the action of field LMXBs. In the very centre of a massive elliptical it seems reasonable to expect an aditional contribution from dynamically formed LMXBs \citep[as in the inner bulge of M31, ][]{VossGilfanov2007}, as well as from LMXBs in globular clusters. Our cooling constant may also be too high for this hot gas regime, although our assumption of constant density will break down.

We can also apply XBF to the other end of the elliptical galaxy mass spectrum, i.e.\ to keeping gas hot in dwarf ellipticals. This is helped by the fact that they are oversized compared to the mass-radius relation used above. For $M_{*}=10^{7} {\rm~M_{\odot}}$ and $r= 1$ kpc then $g \approx 0.3$ per cent. If the appropriate cooling constant is lower by a factor of 10 than our assumed value, then those those galaxies where the XBF luminosity distribution is enhanced by a factor of 10 above the mean (see Section \ref{sec:LMXBflucts}) could keep approximately 25 per cent of their baryonic mass in hot gas through the action of LMXBs. 

\subsection{Star formation}
A similar cooling-to-heating ratio for XBF in young populations at solar metallicity is:
\begin{equation}
\frac{L_{\rm cool}}{L_{\rm XBF}} \approx 5\times10^{-5} \Lambda_{-22} \left( \frac{M_{*}}{\rm M_{\odot}} \right)^{2} \left( \frac{r}{\rm pc} \right)^{-3}  \frac{\dot{M}_{*}}{\rm M_{\odot}~yr^{-1}} .
\end{equation}
We can simplify this by writing the star formation rate ($\dot{M}_{*}$) as $M_{\rm g} \epsilon_{\rm SF} / \tau_{\rm SF}$, where $\epsilon_{\rm SF}$ and $\tau_{\rm SF}$ represent the efficiency and timescale with which gas is turned into stars. This gives:
\begin{equation}
\frac{L_{\rm cool}}{L_{\rm XBF}} \approx 5\times10^{-5} \frac{\tau_{\rm SF} }{\epsilon_{\rm SF}} \Lambda_{-22} \frac{M_{*}}{\rm M_{\odot}} \left( \frac{r}{\rm pc} \right)^{-3}
\end{equation}
i.e. a star formation timescale of $10^{7}$ years and an efficiency of 5 per cent (or $10^{8}$ years and 50 per cent) give a cooling-to-heating ratio of unity.  Note that at low metallicity then not only should XBF become stronger but cooling should also become less efficient.  Furthermore, as discussed in Section \ref{sec:fluctuations}, in many galaxies the XBF input varies from the numbers above simply via stochastic effects.

\end{document}